\documentclass{article}

\PassOptionsToPackage{numbers, compress}{natbib}
\usepackage[final]{nips_2018}

\usepackage[utf8]{inputenc} 
\usepackage[T1]{fontenc}    
\usepackage{hyperref}       
\usepackage{url}            
\usepackage{booktabs}       
\usepackage{amsfonts}       
\usepackage{nicefrac}       
\usepackage{microtype}      
\usepackage{amsmath}

\DeclareMathOperator*{\expectation}{\mathbb{E}}
\title{Deep Nonlinear Non-Gaussian Filtering for Dynamical Systems}

\usepackage{microtype}
\usepackage{graphicx}
\usepackage{wrapfig}
\usepackage{subfigure}
\usepackage{booktabs} 

\author{Arash Mehrjou\\
Department of Empirical Inference\\
Max Planck Institute for Intelligent Systems\\
\texttt{arash.mehrjou@tuebingen.mpg.de}\\
\And
Bernhard Sch\"{o}lkopf\\
Department of Empirical Inference\\
Max Planck Institute for Intelligent Systems\\
\texttt{bs@tuebingen.mpg.de} \\
}

\begin{document}

\maketitle

\begin{abstract}
Filtering is a general name for inferring the states of a dynamical system given observations. The most common filtering approach is Gaussian Filtering (GF) where the distribution of the inferred states is a Gaussian whose mean is an affine function of the observations. There are two restrictions in this model: Gaussianity and Affinity. We propose a model to relax both these assumptions based on recent advances in implicit generative models. Empirical results show that the proposed method gives a significant advantage over GF and nonlinear methods based on fixed nonlinear kernels.
\end{abstract}

\section{Introduction}
Inference in dynamical systems is an standing process in many control systems. We as intelligent agents are constantly inferring the \emph{states} of the nature and systems around us. Observations are often so noisy and unreliable that require us to first infer the underlying states, then make our decisions based on the estimated states. We can use two sources of information to infer the states causing the current observation: (1) The history of our estimate of the previous states (2) The current observation. Fusing these two sources of information to obtain an accurate estimate of the current state of a dynamical system is generally called \emph{filtering}. 

{\bf Dynamical systems---} We assume time-invariant closed-loop dynamical systems described as

\begin{equation}
\left\{
	\begin{array}{ll}
		x_t=f(x_{t-1}, n_t)\\
        y_t=h(x_t, m_t)\\
	\end{array}
    \label{eq:dynamical_system_simple}
\right.
\end{equation}
where the subscript $t$ correspond to the current value and subscript $t-1$ correspond to the values at the moment one step before the current time in discrete setting. This formulation is generic enough for the purposes of this paper; however, the path from the initial formulation to this simplified version can be followed in the Appendix.~\ref{append:dynamical_system}. Obviously, this notation is correct if the system satisfies Markov property. In the above system, $n_t$ and $m_t$ come from some simple noise models. Notice that the simplicity of these noise models is not restrictive because they can be transformed into any complex distribution through the nonlinear functions $f$ and $h$. In a physical system, the first line of $\eqref{eq:dynamical_system_simple}$ describes $p(x_t|x_{t-1})$ as the evolution of the states of the system and the second line describes $p(y_t|x_t)$ as the probabilistic model of the observations (sensors).

{\bf Filtering---} The goal of filtering is to estimate the current state of the system. Assume the subscript $[:t]$ refers to all time instances before the moment $t$ including $t$. At the current moment denoted by subscript $t$, we have seen the history of all observations $y_{:t}=(y_{:t-1}, y_t)$. 
Thus, the inference over the states of a dynamical system can be written as a two-phase process. A \emph{prediction} phase that models our belief about the next state given only the previous observations ($\mathrm{d}x$ is dropped for simplicity throughout the paper):
\begin{equation}
    p(x_t|y_{:t-1})=\int_{x_{t-1}}p(x_t|x_{t-1})p(x_{t-1}|y_{:t-1})
    \label{eq:pred}
\end{equation}

and an \emph{update} phase that modulates our belief about the current state through the Bayes's formula:
\begin{equation}
    p(x_t|y_{:t})=\frac{p(y_t|x_t)p(x_t|y_{:t-1})}{\int_{x_t}p(y_t|x_t)p(x_t|y_{:t-1})}
    \label{eq:update}
\end{equation}
Kalman derived the closed-from solution for a linear process $\dot{X}=AX+Bv$ and Gaussian noise $v\sim\mathcal{N}(0, I)$~\citep{kalman1960new}. The problem is though very difficult for nonlinear process $f$ and sensor model $h$ unless in very restricted cases~\citep{benevs1981exact, daum1986exact}. The actual goal of filtering is often not computing the posterior distribution of states. Instead, the goal is computing some expectation $\expectation[g(x_t)]$ of a function $g$ of the current state with respect to $p(x_t|y_{:t})$ or $p(x_t, y_t|y_{:t-1})$. The former results in an intractable integral whose computation scales exponentially with the state dimension $\dim(x)$~\citep{wuethrich2016new}. However, computing the latter scales linearly with $\dim(x)$. Even though the integral with respect to the probability measure $p(x_t, y_t|y_{:t-1})$ is computationally feasible, approximating the probability distribution itself is difficult for high dimensional states and observations. This problem has been approached by various methods including Unscented Kalman Filter (UKF)~\citep{julier1997new}, Extended Kalman Filter (EKF)~\citep{sorenson1985kalman} and Particle Filter (PF)~\citep{gordon1993novel} where the first two are parametric and the last one is non-parametric. Most of the parametric methods adopt a variational approach and approximate $p(x_t, y_t|y_{:t-1})$ by $q(x_t, y_t|y_{:t-1})$ that belongs to a parametric hypothesis space. The assumed form for $q$ must be in a way that eases the conditioning on $y_t$ which is readily possible for a Gaussian $q$. Nonetheless, a Gaussian distribution is not a realistic assumption for $p$ unless in very limited applications. In this paper, we propose an easily trainable and highly expressive variational distribution and an efficient method to learn its parameters.

\section{Gaussian Filtering}
In common filtering applications, what we usually care about is an expectation of the following form:

\begin{align}
    \expectation[g(x_t, y_t)] = &\int_{x_t, y_t} g(x_t, y_t)p(x_t, y_t|y_{:t-1}) = \label{eq:expectation_pre_plugging}
    &\int_{x_t, m_t} g(x_t, h(x_t, m_t)) p(m_t)p(x_t|y_{:t-1})
\end{align}
where the right-hand integral is derived by plugging in the observation model of ~\eqref{eq:dynamical_system_simple} in~\eqref{eq:expectation_pre_plugging}. This integral is computable by Monte Carlo methods when the distribution $p(x_t|y_{:t-1})$ can be sampled efficiently and noise has a simple model $p(m_t)$.  As an special case, integrals with respect to $p(x_t|y_{:t-1})$ can be computed efficiently as well. However, it requires $p(x_t|y_t)$ to be easily computable from $p(x_t, y_t)$ which is not the case for most distributions except very simple ones such as Gaussians. 
To ease the presentation, let's focus only on the prediction step~\eqref{eq:pred} to compute $p(x_t|y_{:t-1})$. The history of observations $y_{:t-1}$ is implicit in the model. Thus, we drop the indices and represent $x_t$ by $x$ and $y_t$ by $y$. For example, $p(x_t|y_{:t})=p(x_t|y_t, y_{:t-1})$ is simply represented by $p(x|y)$.
As mentioned in the previous section, filtering tries to find a good approximation to $p(x, y)$ and ultimately $p(x|y)$. This process is carried out by first approximating $p(x,y)$ by $q(x,y)$, then computing $q(x|y)$ given $q(x,y)$. The distribution $q(x,y)$ is often chosen from a hypothesis space with limited capacity. For a Gaussian hypothesis set, we have 
\begin{align}
    q(x,y) &= \mathcal{N}\left(
\left(\!
    \begin{array}{c}
      x \\
      y
    \end{array}
  \!\right)\middle| \left(\!
    \begin{array}{c}
      \mu_x \\
      \mu_y
    \end{array}
  \!\right), \left(\begin{matrix}
\Sigma_{xx}&\Sigma_{xy} \\ \Sigma_{yx}&\Sigma_{yy}
\end{matrix} \right)
  \right)\\
  q(x|y) &= \mathcal{N}(x|\underline{\mu_x + \Sigma_{xy}\Sigma_{yy}^{-1}(y-\mu_y)}, \Sigma_{xx}-\Sigma_{xy}\Sigma_{yy}^{-1}\Sigma_{xy}^T) \label{eq:GF_posterior}
\end{align}

which is in general called Gaussian Filter (GF). There are two obvious limitations in this framework: First, the posterior distribution~\eqref{eq:GF_posterior} is Gaussian. Second, the mean of the posterior distribution of states which is underlined in~\eqref{eq:GF_posterior} is an affine function of the the observations $y$. In the next section, we relax both these assumptions.

\begin{figure}[t!]
	\centering

\subfigure[State-Observation evolution]{
\includegraphics[width=0.40\linewidth]{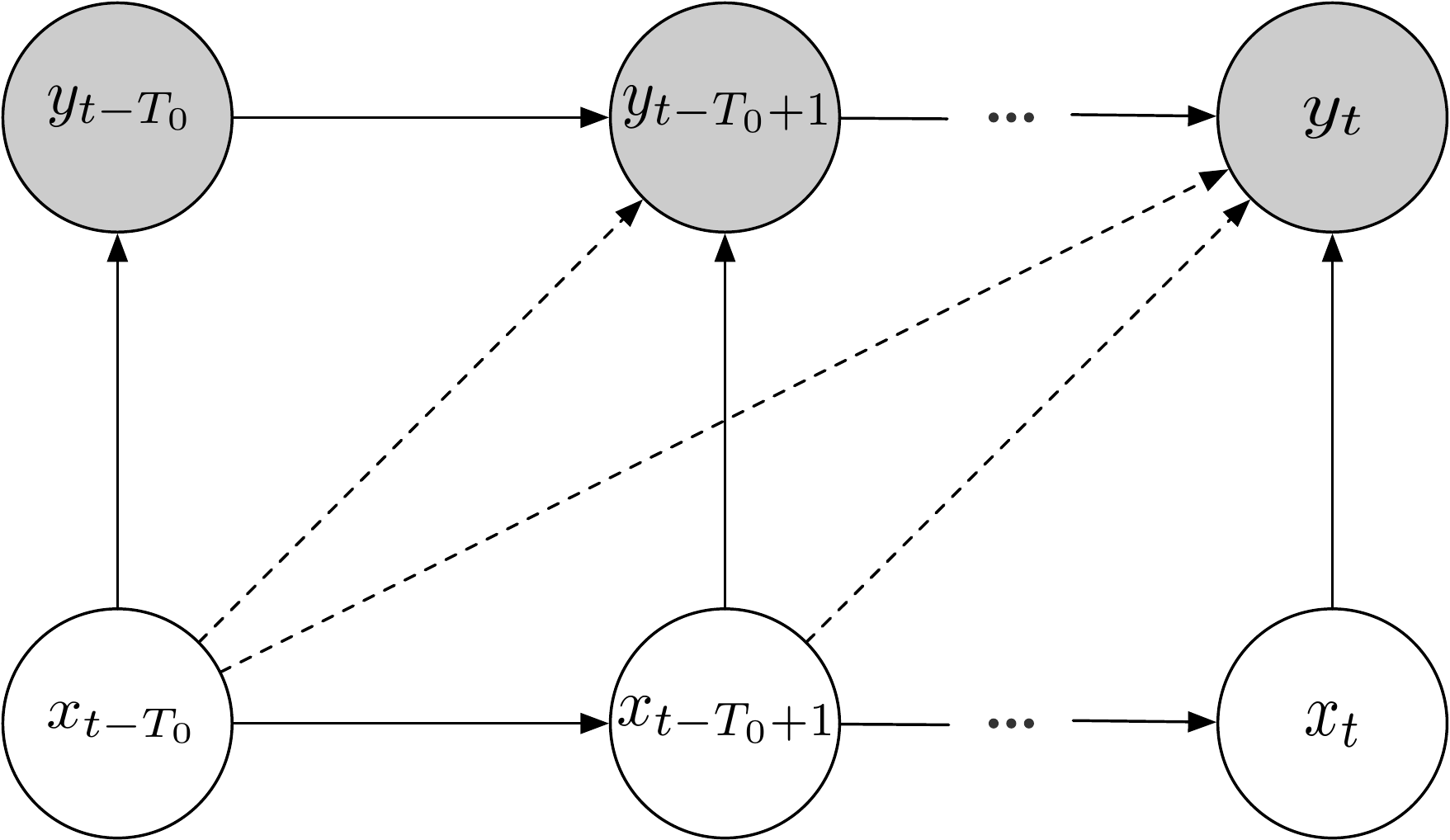}}
\label{fig:architecture}
\hspace{1cm}
\subfigure[The architecture for sampling from the posterior]{
\includegraphics[width=0.50\linewidth]{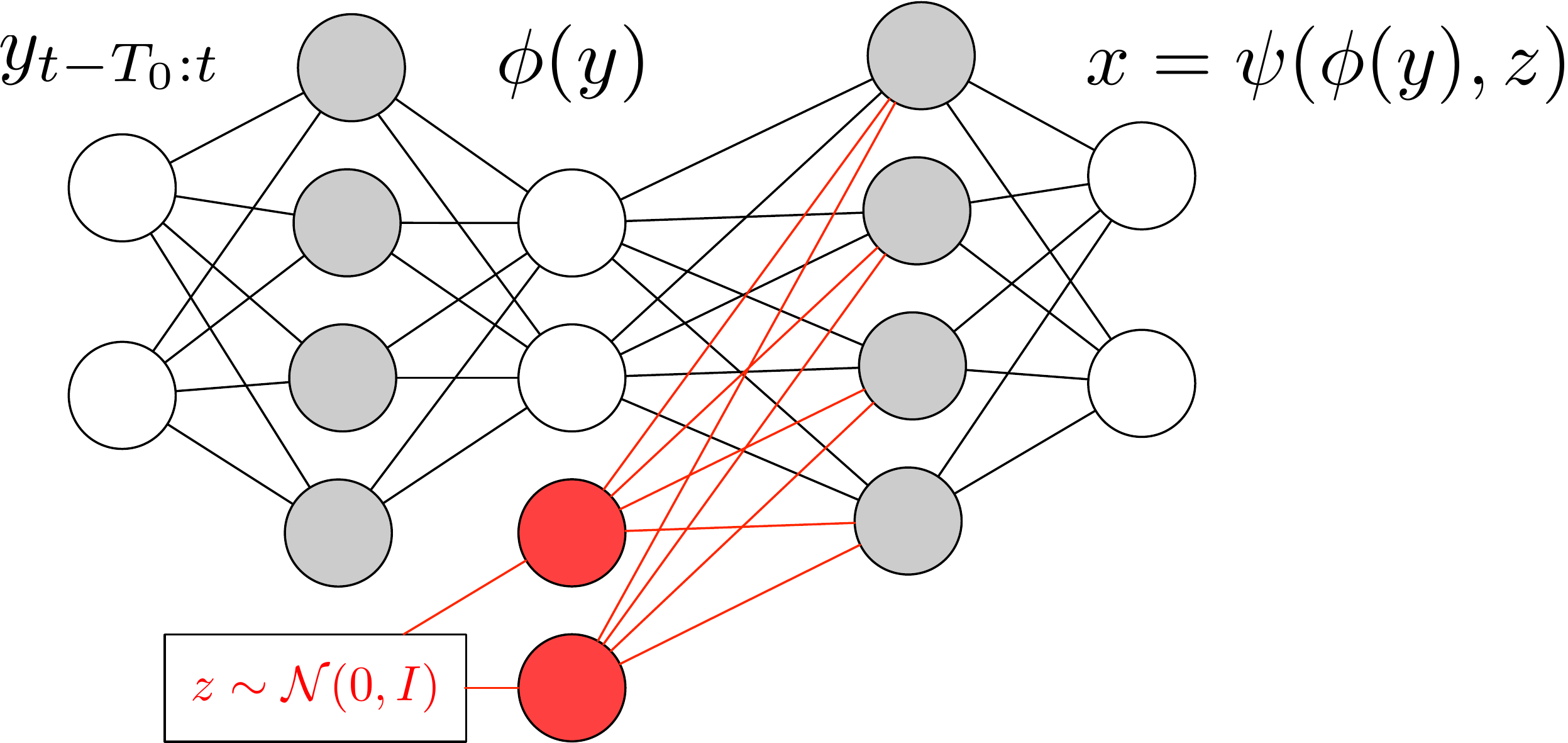}}
\caption{\small (a) Solid lines show how observations are generated by the evolution of states in Markovian setting. Dashed lines show the non-Markovian setting. (b) The observations of $T_0$ previous timesteps is fed to the network and are transformed to the nonlinear features. The features are concatenated with samples from an external source of noise ($z$) and passed through a nonlinear function whose output is supposed to match the samples from the posterior distribution of states given observations of the last $T_0$ timesteps.}
\label{fig:sampling_sphere}
\end{figure}

\section{Nonlinear Non-Gaussian Filtering}

We take a nonlinear approach and directly approximate the conditional distribution $p(x|y)$ of~\eqref{eq:update} by Multilayer Perceptron (MLP) as a \emph{universal function approximator}~\citep{hornik1991approximation}. In this formulation, $q(x|y)=\mathcal{D}(x|\phi(y))$ where $\phi$ is a nonlinear function of $y$. Moreover, $\mathcal{D}$ can be any complex distribution over $x$ belonging to $n$-dimensional state space. We do not compute $\mathcal{D}$ directly. Rather, we generate samples $x_i$ such that $x_i\sim\mathcal{D}$.  In analogy with kernel machines, we call $\phi:\mathbb{R}^m\to \mathbb{R}^M$ a feature extractor that transforms the measurements by a nonlinear function from the $m$-dimensional \emph{sensor} space to the $M$-dimensional feature space. Let's assume $\phi$ is parameterised by an MLP as $\phi(y;\theta_\phi)$. This is after all a deterministic mapping and lacks  the required stochasticity. Therefore, we provide the stochastic \emph{fuel} to $q(x|y)$ by passing samples $z\sim\mathcal{N}(0, I)$ alongside the extracted features $\phi(y;\theta_\phi)$ through a secondary parameterized function $\psi(z, \phi(y;\theta_\phi);\theta_\psi)$. Back-propagation is then used to perturb the parameters $\theta_\phi$ and $\theta_\psi$ to make the output of $\psi$ close to the samples from $p(x|y)$. The overall architecture partly inspired by~\citep{bouchacourt2016disco} is shown in Fig.~\ref{fig:architecture}(b). The dashed arrows in Fig.~\ref{fig:architecture}(a) suggests the possibility of weakening the Markovian assumption of~\eqref{eq:dynamical_system_simple} such that distant states in the past can influence the current observation. Despite the difficulty of filtering for non-Markovian systems in other methods~\citep{wuethrich2016new}, the proposed method can take care of it simply by feeding more observations from the past in the network as depicted in Fig.~\ref{fig:architecture}(b).

{\bf Learning the state posterior--- }The proposed method is expected to accurately capture the posterior distribution $p(x|y)$ by $q(x|y)$ where $q(x|y)$ is much more flexible than Gaussian. We define the loss $\ell:\mathcal{X}\times\mathcal{X}\to \mathbb{R}^+\cup \{0 \}$ as a simple Euclidean distance $\ell(x,x')=\lVert x-x'\rVert_2^2$ for $x\sim p(x|y)$ and $x'\sim q(x|y)$. Since MLP can theoretically capture arbitrarily complex functions~\citep{hornik1991approximation}, we move on one step further and make $q(x(t)|y(t))$ a function of not only $y(t)$ but also a few previous observations of the system that turns the implicit distribution into $q(x|y_{t-T_0:t})$ where $T_0$ is the approximate time interval in the past through which the observations are informative about the current hidden state of the dynamical system. Inspired by~\citep{bouchacourt2016disco}, given any non-negative symmetric loss function $l(x,x')$ for $(x, x')\in \mathcal{X}\times\mathcal{X}$, we define \emph{diversity coefficient} as
\begin{equation}
    \mathrm{\Delta}(p,q)=\expectation_{y\sim p(y)}[\hspace{-1.55cm}\expectation_{\substack{\\\hspace{1.55cm}x^\prime\sim q(x^\prime|y_{t-T_0:t})\\\hspace{1.55cm}x\sim p(x|y_{t-T_0:t})}}\hspace{-1.55cm}[\ell(x,x^\prime)]]
\end{equation}

On the other hand, due to the uncertainty in the posterior, we know that $q(x^\prime|y_{t_{t-T_0:t}})$ should not collapse to an extremely low entropy distribution. To encourage the implicitly estimated posterior to have higher entropy, we add a diversity encouraging term to the loss function where similarity among the samples from the same distribution acts as repulsive forces. Hence, the overall loss function becomes
\begin{equation}
\label{eq:loss}
    \mathcal{L}_\ell(q) =  \mathrm { \Delta } _ { \ell } (p,q) - \lambda \mathrm { \Delta } _ { \ell } (q,q)
\end{equation}
where $\lambda$ is a hyper-parameter that roughly controls the empirical entropy of the samples generated from the implicit variational posterior $x\sim q(x|y)$.

{\bf Optimization--- } In practice, the loss function~\eqref{eq:loss} is approximated empirically by the sums:
\begin{align}
    \hat{\mathrm { \Delta }} _ { \ell} (p,q_{\theta_\phi,\theta_\psi})&=\frac{1}{N}\sum_{n=1}^N\frac{1}{K}\sum_{k=1}^K l(x_n,\psi(\phi(y_n;\theta_\phi),z_k;\theta_\psi))\nonumber\\
    \hat{\mathrm { \Delta }} _ { \ell} (q_{\theta_\phi,\theta_\psi},q_{\theta_\phi,\theta_\psi})&=\frac{1}{N}\sum_{n=1}^N\frac{1}{K(K-1)}\sum_{k=1,k'\neq k}^K l(\psi(\phi(y_n;\theta_\phi),z_k;\theta_\psi), \psi(\phi(y_n;\theta_\phi),z_{k'};\theta_\psi))\nonumber
\end{align}
and the loss function $\hat{\mathcal{L}}(\theta_\phi,\theta_\psi)=\hat{\mathrm { \Delta }} _ { \ell} (p,q_{\theta_\phi,\theta_\psi})+\hat{\mathrm { \Delta }} _ { \ell} (q_{\theta_\phi,\theta_\psi},q_{\theta_\phi,\theta_\psi})$ is optimized with respect to $\{\theta_\phi,\theta_\psi\}$ by gradient descent. Notice that, for each training example $(x_n,y_n)$ from the training set, we need to sample $K$ values of external noise $z$. The greater $K$ results in faster convergence.

\begin{figure}[t!]
\vspace{-1 cm}
	\centering
\subfigure[Gaussian filter]{
\includegraphics[width=0.47\linewidth]{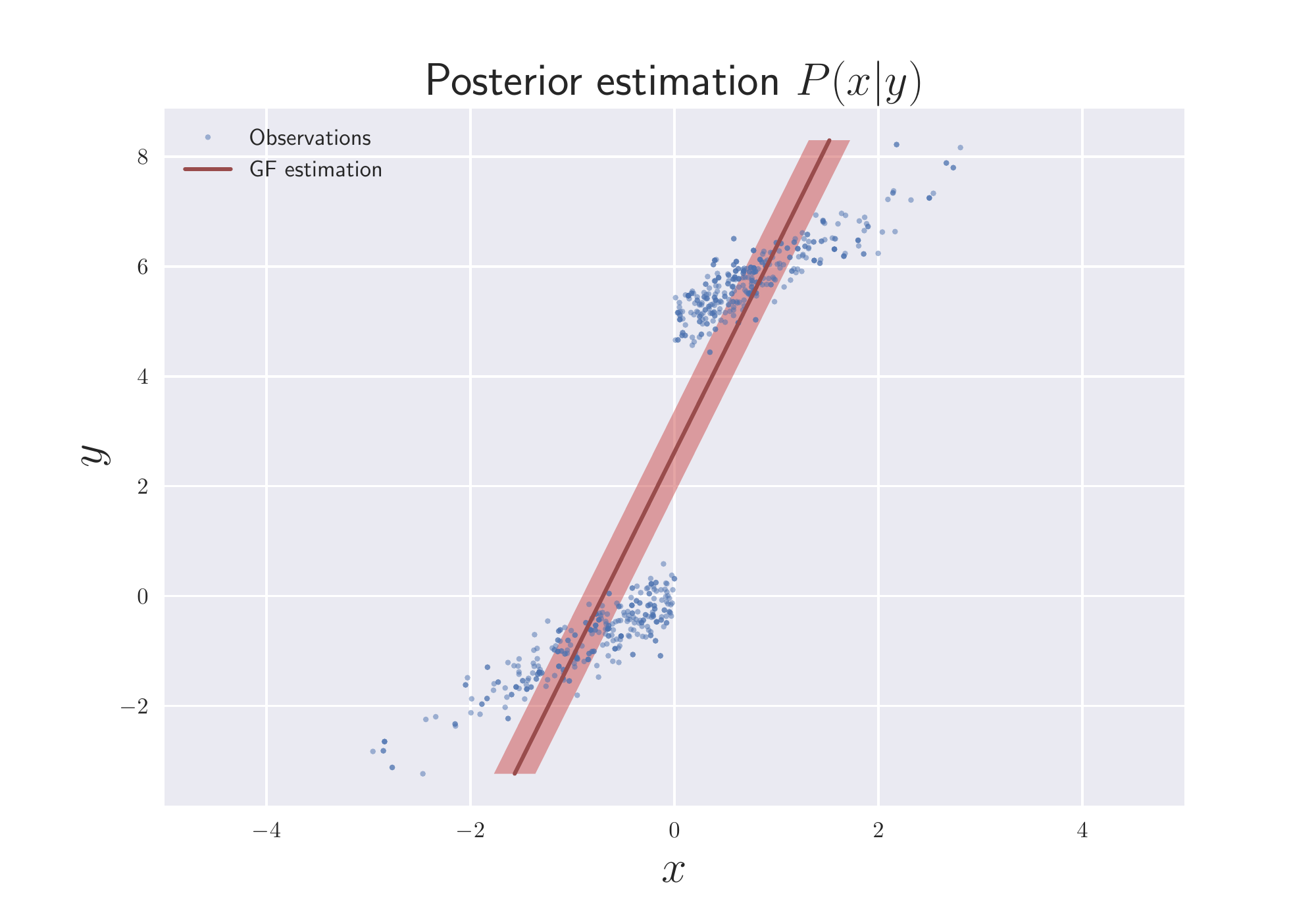}}
\subfigure[Proposed method]{
\includegraphics[width=0.47\linewidth]{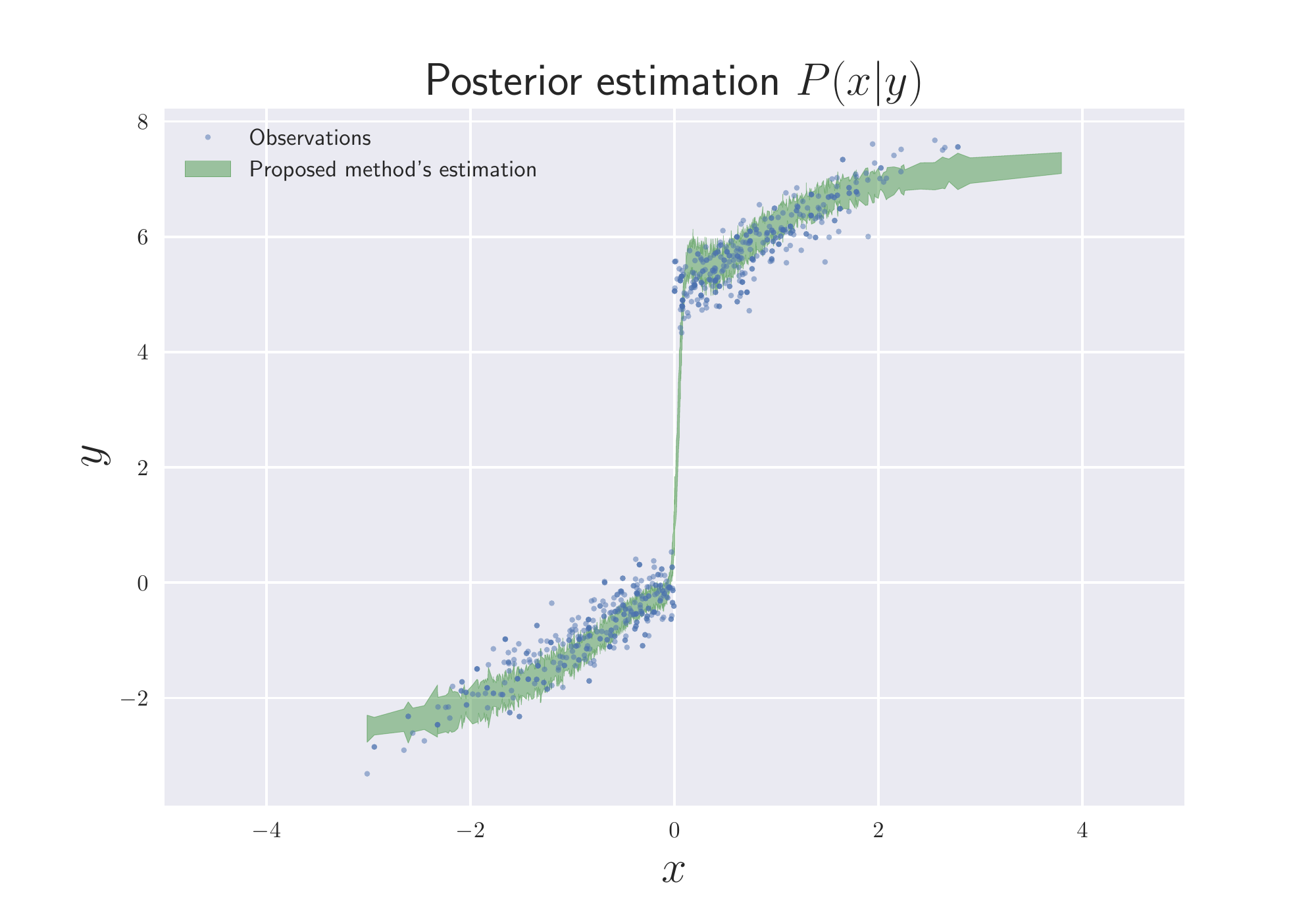}}\\
\vspace{-4mm}
\subfigure[Nonlinear Gaussian Filter (degree=3)]{
\includegraphics[width=0.47\linewidth]{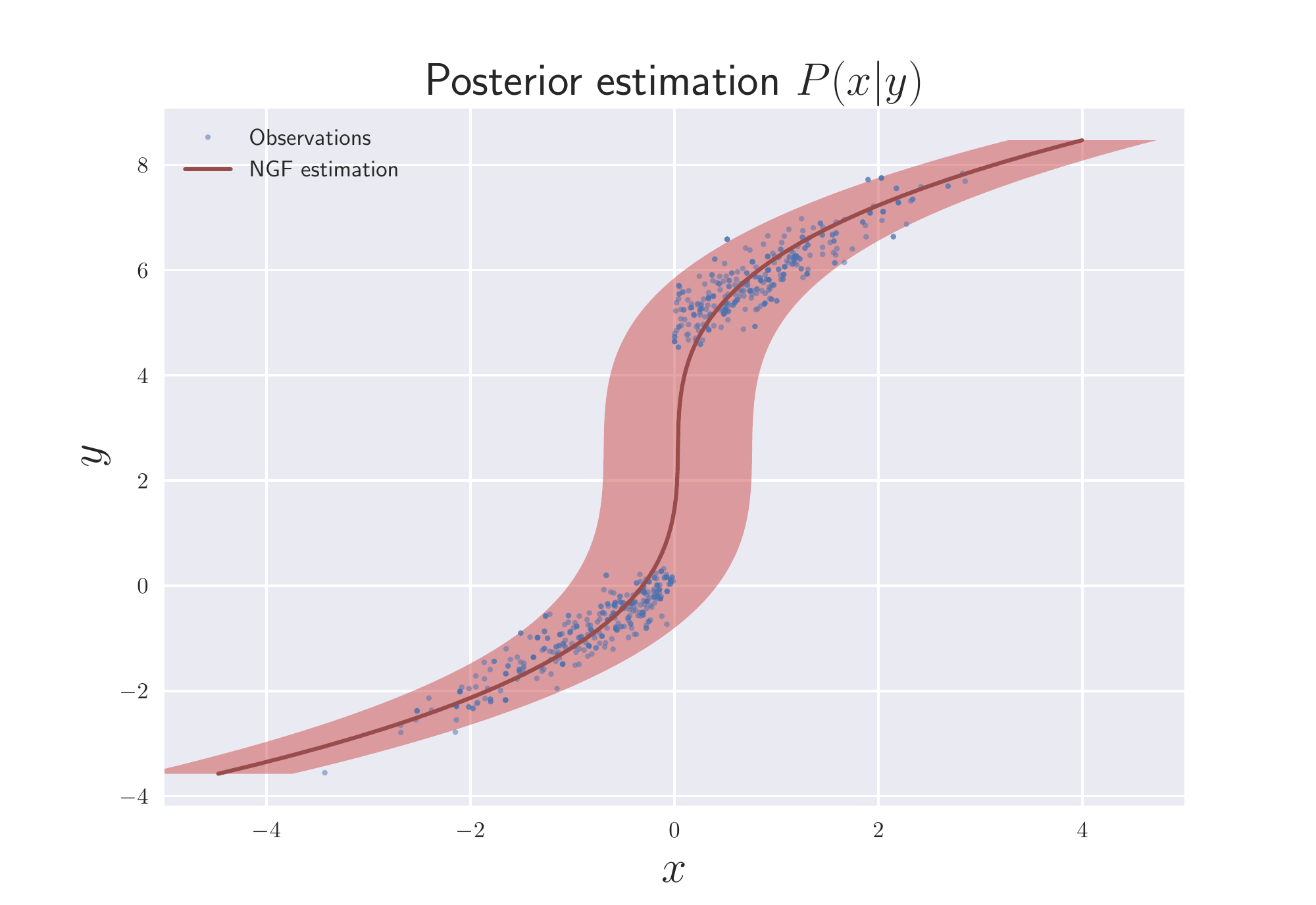}}
\subfigure[Nonlinear Gaussian Filter (degree=7)]{
\includegraphics[width=0.47\linewidth]{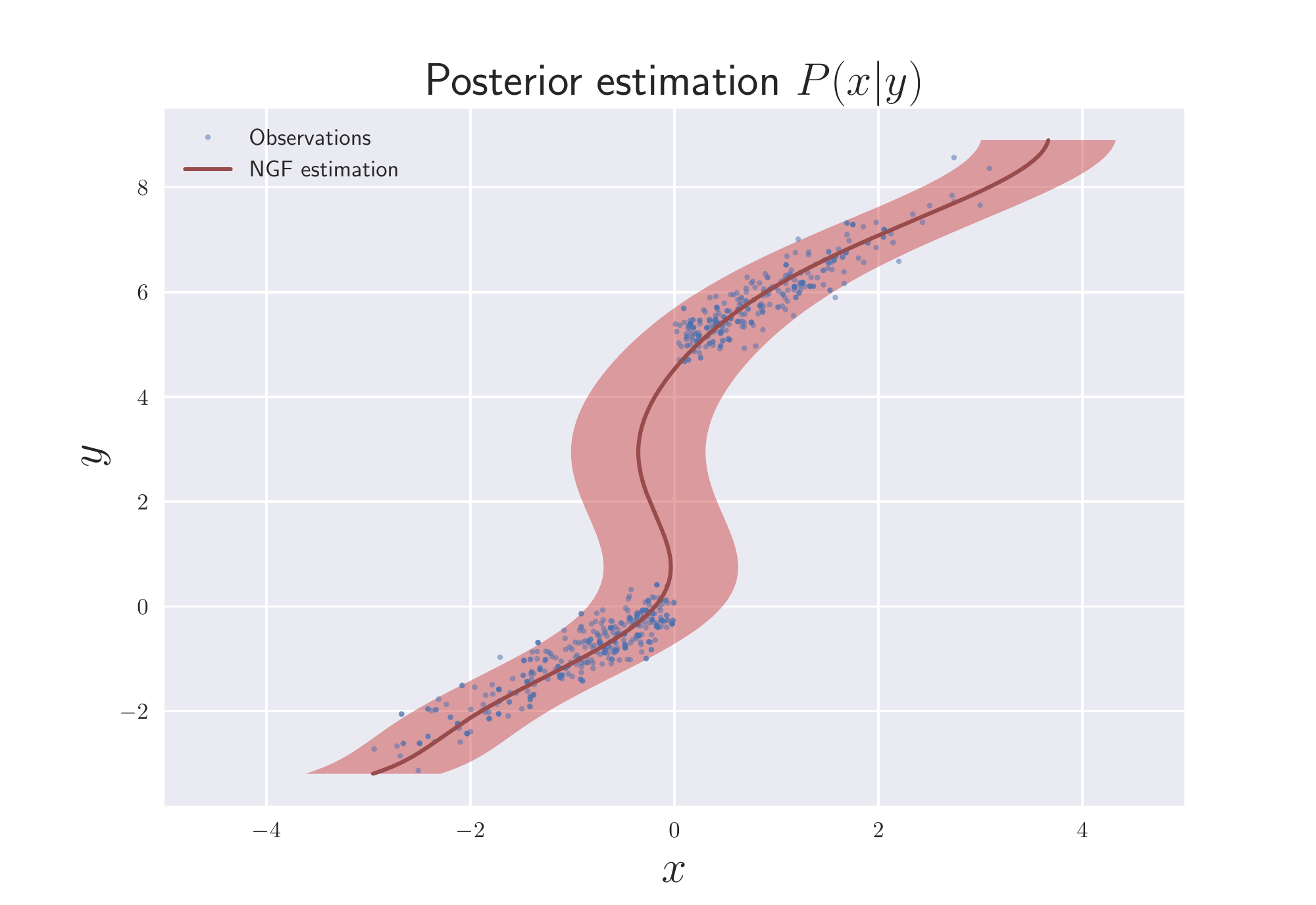}}
\caption{\small Estimated posterior by different methods. Shaded area shows the standard deviation around the mean. Notice that the observation is on the vertical axis, so the uncertainty is the width of the shaded area along the horizontal axis. As can be seen, GF(a) is quite inaccurate since the posterior is Gaussian and its mean is only an affine function of the observation. NGF(c) with polynomial nonlinearity with degree 3 fits the posterior better than GF but it needs a good choice of nonlinearity to give an acceptable result. Otherwise, the estimated posterior can be too simple or too complex(d). Moreover, this method becomes very uncertain around $x=0$ which means it can gives many different values for $x$ when $y\in[0,4]$. The proposed method shows a good performance which is caused by two trainable nonlinearity. Roughly speaking, $\phi$ approximates the mean and $\psi$ approximates the variance of $p(x|y)$ corresponding to each observation $y$.}
\label{fig:results}
\end{figure}

{\bf Experiments--- }We compare the performance of the proposed method with Gaussian Filter (GF) and also the Nonlinear Gaussian Filter (NGF) ~\citep{wuethrich2016new} where a fixed nonlinear feature extractor is used to transform sensor measurements to the feature space. The system is described by $g(x_{t-1}, n_t) = x_{t-1} + n_t$, $h(x_t,m_t)=x_t+m_t+5H(x_t)$ and $p(x_{t-1})=\mathcal{N}(x_{t-1}|0,5)$ where $H(.)$ is the Heaviside step function. See Fig.~\ref{fig:results} and its caption for description and Appendix~\eqref{append:desc_experiment} for more details.

{\bf Conclusion--- } We proposed a method that learns to filter the states of dynamical systems given previous values of sensor measurements and states. It benefits from the flexibility of MLPs to deal with major limitations of other methods such as linearity, Gaussianity, and Markovian assumption in a simple unified way. Notice that the method generates samples from the posterior which is enough for computing the integral~\eqref{eq:update}. The method however cannot be used in possible applications where the evaluation of $p(x|y)$ is required and also where some samples from (observations, states) pair are not available to accomplish the learning phase.

\newpage
\bibliographystyle{unsrt}      
\bibliography{main}

\newpage
\section*{Appendices}
\appendix 
\section{Description of dynamical systems}
\label{append:dynamical_system}
We assume the generic formulation of dynamical systems as follows
\begin{equation}
\left\{
	\begin{array}{ll}
		\dot{x}(t)=f(x(t), u(t), t)\\
        y(t)=h(x(t), u(t), t)\\
	\end{array}
    \label{eq:dynamical_system_general}
\right.
\end{equation}
As a simplifying assumption, we ignore the explicit dependence on time for now. Moreover, we assume the system is closed loop, i.e., $u(t)$ is designed by state feedback to be a function of states as $u(x(t))$. By considering the system as time invariant and discrete (which is the case in practice where reading the sensors and issuing control signals are performed by digital systems), we denote the current moment by $t$ and one timestep before it by $t-1$. Therefore, the description of the dynamics $f$ and the observation model $h$ are simplified to~\eqref{eq:dynamical_system_simple}.

\section{Details on the experiment}
\label{append:desc_experiment}
The experiment was performed for the following nonlinear stochastic dynamical system proposed in~\citep{wuethrich2016new}:
\begin{equation}
\left\{
	\begin{array}{ll}
		\dot{x}_t= x_{t-1} + n_t\\
        y_t = x_t+m_t+5H(x_t)\\
	\end{array}
\right.
\end{equation}
where the subscripts have the meaning that was described earlier for~\eqref{eq:dynamical_system_simple}. The state and observation noises are both Gaussian with variances $0.1$ and $0.3$ accordingly. The value of $\lambda$ in the loss function~\eqref{eq:loss} is set to $1$. However, we obtaine comparable results for a range of $\lambda$ values between $0.7$ to $2.5$. The training set is generated by running the dynamical system for $1000$ time instances starting from a random starting state $x_0\sim\mathcal{N}(0,1)$ . The generated training set is then used to train the loss function~\eqref{eq:loss} by Adam optimizer~\citep{kingma2014adam} with learning rate $0.005$, decay rate $0.95$ per every $100$ iterations and batch size 20. Training the networks has been continued for approximately 3000 iterations until the value of the parameters converge. Notice that the shaded area shows the standard deviation around the mean. In GF and NGF, standard deviation has a closed-form formula. In the proposed method, since the posterior distribution is implicit and we only have access to samples generated from the approximated posterior, the variance is computed empirically using the generated samples from $q(x|y)$ and plotted to show that the diversity in the generated samples matches the diversity of the actual posterior distribution $p(x|y)$.

{\bf Network architecture--- } We used almost the same network architecture for both $\phi$ and $\psi$ functions. It consists of two hidden layers each with $128$ neurons and tanh nonlinearity. These layers are followed by a linear output layer. The only difference between the networks realizing $\phi$ and $\psi$ is that the former has an output layer with $10$ neurons meaning that the feature space to which the sensor measurements are transformed is $10$-dimensional. On the other hand, the function $\psi$ obviously has the same output dimension as the dimension of the state $x$. We did experiments with several other dynamical systems with different state/sensor dimensions and constantly observed improvement over GF and NGF.

\end{document}